\documentclass[aps,prl,graphicx,9pt,twocolumn]{revtex4-2}
\usepackage{amsmath}
\usepackage{amscd}
\usepackage{graphicx}
\usepackage{cases}
\usepackage{multirow}
\usepackage{float}
\usepackage{subfigure}
\usepackage{epstopdf}
\usepackage{appendix}
\usepackage{diagbox}
\usepackage{makecell}
\usepackage{blindtext}
\usepackage{amsmath}
\usepackage{tabularx}

\usepackage[colorlinks,
linkcolor=blue,
urlcolor=blue,
anchorcolor=blue,
citecolor=blue]{hyperref}
\begin{document}
\title[Short Title]{Demonstration of dynamical control of three-level open systems with a superconducting qutrit}
\author{Ri-Hua Zheng }
\thanks{These two authors contributed equally to this work.}
\author{Wen Ning}
\thanks{These two authors contributed equally to this work.}
\author{Zhen-Biao Yang}\email{zbyang@fzu.edu.cn}
\author{Yan Xia}\email{xia-208@163.com}
\author{Shi-Biao Zheng}\email{t96034@fzu.edu.cn}
\affiliation{Fujian Key Laboratory of Quantum Information and Quantum Optics, College of Physics and Information Engineering, Fuzhou University, Fuzhou, Fujian 350108, China}
\begin{abstract}
We propose a method for the dynamical control in three-level open systems and realize it in the experiment with a superconducting qutrit. Our work demonstrates that in the Markovian environment for a relatively long time (3 $\mu$s), the systemic populations or coherence can still strictly follow the preset evolution paths. This is the first experiment for precisely controlling the Markovian dynamics of three-level open systems, providing a solid foundation for the future realization of dynamical control in multiple open systems. An instant application of the techniques demonstrated in this experiment is to stabilize the energy of quantum batteries.
\end{abstract}	
\maketitle
\textit{Introduction}.--The control of quantum systems is always an extremely interesting and essential topic among diverse types of quantum information tasks \citep{QC-STIRAP,QC-molecular}. With the gradual development of experimental techniques, decoherence, mainly resulted from energy relaxation and dephasing, has become a primary barrier lying on the way to experimentally implement the quantum control. To clear this barrier, the researchers choose to take measures to shorten the evolution time and have achieved remarkable successes in the experiments, such as, preparing 20-qubit entanglement \citep{Song-20bitcat,Omran-20bitcat} and simulating quantum walk on 62-qubit processor \citep{zhu-62bit}. Meanwhile, some researchers attempt on constructing qubits that are insensitive to decoherence \citep{zhu-T1-40us,zhu-62bit,zuchonzhi,Qiskit}.  So far, the relaxation time $T_1$ and dephasing time $T_2$ of superconducting qubit have both been extended to 200 $\mu$s \citep{Qiskit}. Indeed, the methods of shortening the evolution time and making more perfect qubits naturally reduce the impact of decoherence rather than avoiding it, which still process quantified errors. For realizing the leap of quantum error rates from $10^{-3}$ to $10^{-15}$ \citep{PRAe-15,-3to-15} and further accomplishing near-term quantum works, it is necessary to explore the control of open quantum systems, tailoring the decoherence as one of the effective elements involved in the quantum control.

The earlier open-system quantum control theory draws supports from a kind of nuclear magnetic resonance technology, called spin-echo \citep{Echo}, in which a $\pi$ pulse is inserted midway in the evolution time to increase $T_2$. However, this inserted $\pi$ pulse will change the systemic dynamics. Several spin-echo-like techniques, such as bang-bang control \citep{bangbang-PRA}, dynamical decoupling technique \citep{DD-PRL}, and parity-kicks \citep{kicks-PRA}, process the same problems with the destruction of dynamics. So these works generally served as techniques for storing coherence of quantum systems \citep{DDapply1,DDapply2,DDapply3,DDapply4,6hours,Guo-longT2}. 
Researchers alternatively develop an open-system adiabatic theorem \citep{open-adiabatic} to fully control the dynamics. This method utilizes super operators to describe the open systems and can approximately predict the density matrix at each moment. Since the adiabatic theorem needs a long evolution time to guarantee the adiabatic approximation, researchers further try to build a shortcut to adiabaticity (STA) (similar to that in the closed systems \citep{STA-Rice,STA-Berry,STA-Chen1,STA-Chen2,STA-RE1,STA-RE2,STA-yehong1,STA-yehong2,rihua1,rihua2,yihao1,yihao2,su1,su2}). 

By virtue of super operators, Sarandy $et$ $al$. \citep{open-invariant} have proposed a concept about open-system dynamical invariants to build a STA in open systems. They constructed some 4-dimensional invariants for engineering two-level open systems, nevertheless, with several simplifications, such as, either considering energy relaxation or dephasing. Finding proper invariants will be complicated when someone simultaneously takes energy relaxation and dephasing into account, let alone expansion to three-level open systems with 9-dimensional invariants. Therefore other control methods are eagerly in need of discovery.

Recently, Medina and Semi\~{a}o \citep{pop-control2} have directly set the density operator with time-dependent parameters and then submitted them to the Markovian master equation \citep{master-eq}, which gives appropriate functions of pulses to control the populations in two-level open systems. The error rates of populations control in Ref. \citep{pop-control2} under serious decoherence are almost zero, an exciting result. Some follow-up studies \citep{ran-PRA,ran-OL} have made a lot of additions to the application scenarios, but still remaining in two-level open systems. With the general trend of quantum tasks towards multiplication \citep{multicodePRL}, it is of great significance to explore the dynamics of three (or more)-level open systems. 

In this letter, we propose a method for the dynamical control in three-level open systems, and demonstrate it with a superconducting circuit platform \citep{Barends-Nature,Kelly-Nature,caoNC,ningPRL,Xukai-optica,YangNPJ}. Two time-dependent parameters are presupposed in the Markovian master equation to singly control the populations or coherence, and then the analytical functions of driving pulses are given to complete desired dynamical control. For proving the correctness of this method, we apply the designed driving pulses to a superconducting Xmon qutrit and measure populations of excited states $|1\rangle$ and $|2\rangle$, or coherence between ground and excited states. The experimental results are in good agreement with the theoretical design for the dynamical control. This is the first experiment to precisely control the Markovian dynamics of three-level open systems, where the decoherence is effectively dominated. In addition, an interesting freezing phenomenon of populations is observed. That is, in the late stages of evolutions, the populations stabilize at specific values for a relatively long time ($>$1.8 $\mu s$) in the presence of decoherence. An instant application scenario of this freezing phenomenon is the quantum battery (QB) \citep{QB-initial,QB-PRL1,QB-PRL2,QB-PRL3,QB-NJP,QB-PRB,QB-PRA1,QB-PRA2,QB-PRA3,QB-PRE,QB-EPL}. One can steadily lock the energy of QBs for a comparatively long time ($>$1.2 $\mu s$), resisting the relaxation of excitation, and therefore building prototypes of QBs with more stable energy. 

%Like Ref. \citep{pop-control2,ran-PRA,ran-OL}, there are some unacceptable dynamics because of decoherence. Ergo, we also provide precise restricted conditions to depict such a phenomenon, increasing physical stringency of this method. Numerical simulation indicates that the present method grantees almost 100\% fidelity for population transformation when considering serious decoherence.

\textit{Method}.--Assume that a three-level system consists of bases $|0\rangle$, $|1\rangle$, and $|2\rangle$. The systemic Hamiltonian in the interaction picture is ($\hbar=1$ hereafter) $H_i(t)=\Omega_{01}(t)|0\rangle\langle 1|+\Omega_{12}(t)|1\rangle\langle 2|+{\rm H.c.}$, where $\Omega_{01}(t)$ and $\Omega_{12}(t)$ respectively represent the Rabi frequencies of the designed driving pulses for controlling transitions $|0\rangle \leftrightarrow |1\rangle$ and $|1\rangle \leftrightarrow |2\rangle$. We assume the density matrix of this three-level system being [basis order \{$|2\rangle$,$|1\rangle$,$|0\rangle$\} and dropping $(t)$ henceforth]
%why such like? 
\begin{eqnarray}\label{rho}
\rho=\left(
\begin{array}{ccc}
f_2 & -ih_1 & h_2 \\
ih_1 & f_1 & -ih_3 \\
h_2 & ih_3 & 1-f_1-f_2 \\
\end{array}
\right),
\end{eqnarray}
in which  $f_1$, $f_2$, $h_1$, $h_2$, and $h_3$ are all time-dependent real functions. Here  $f_1$ and $f_2$ describe the populations of states $|1\rangle$ and $|2\rangle$, respectively. Additionally, $h_1$, $h_2$, and $h_3$ describe the coherence of this three-level system.
Equation \ref{rho} (5 degrees of freedom) is not the most generalized case for a three-level density operator (8 degrees of freedom), however sufficient as a demonstration. 
The Markovian master equation \citep{master-eq} here is 
\begin{eqnarray}\label{mastereq}
\dot{\rho}&=&-i[H_i,\rho]+\sum_{j=1}^{4}\big[L_j \rho L_j^\dag - \frac{1}{2}(L_j^\dag L_j \rho+\rho L_j^\dag L_j)\big],
\end{eqnarray}
where Lindblad operators are $L_{k}=\sqrt{\gamma_k}|k\rangle \langle k|$ and $L_{k+2}=\sqrt{\Gamma_k}|k-1\rangle \langle k|$, with $\Gamma_k$ and $\gamma_k$ being the rates of energy relaxation and dephasing of state $|k\rangle$, respectively ($k=1,2$).
By utilizing Eqs. (\ref{rho}, \ref{mastereq}), we can derive the Rabi frequencies to control the dynamics of this three-level system, yielding
\begin{small}
\begin{subequations} \label{omega}
\begin{align}
\Omega_{01}=&\frac{f_1\Gamma_1+\dot{f}_1+\dot{f}_2}{2h_3}, \\ \Omega_{12}=&\frac{f_2\Gamma_2+\dot{f}_2}{2h_1}.
\end{align}
\end{subequations}
\end{small}
It is noteworthy that there are three constraint equations
\begin{small}
\begin{subequations} \label{consteq}
\begin{align}
\dot{h}_1&=-\frac{1}{2}(\gamma_1+\gamma_2+\Gamma_1+\Gamma_2)h_1+(f_1-f_2) \Omega_{12}-h_2\Omega_{01},  \\
\dot{h}_2&=-\frac{1}{2}(\gamma_2+\Gamma_2)h_2-h_3\Omega_{12}+h_1\Omega_{01}, \\
\dot{h}_3&=-\frac{1}{2}(\gamma_1+\Gamma_1)h_3+h_2\Omega_{12}-(-1+f_2+2f_1)\Omega_{01}.
\end{align}
\end{subequations}
\end{small}
Up to now, the populations or coherence of the three-level system can be effectively controlled by the interaction Hamiltonian $H_i$, i.e., by Rabi frequencies $\Omega_{01}$ and $\Omega_{12}$. A specific example is, an evolution from the ground state $|0\rangle$ to a final state with  populations respectively being $P_0(t_f)$, $P_1(t_f)$, and $P_2(t_f)$ in states $|0\rangle$, $|1\rangle$, and $|2\rangle$, can be achieved by adjusting $f_1=fP_1(t_f)$ and $f_2=fP_2(t_f)$. The intermediate function $f$ changing from 0 to 1 in the time interval $[0,t_f]$ can be set as $f=[1+e^{-a(t-t_f/2)}]^{-1}$ \citep{STA-RE1,STA-RE2,pop-control2}, with $a=50/t_f$ determining the gradient of the transformation. By submitting $f_1$ and $f_2$ into Eqs. (\ref{omega}, \ref{consteq}), one can obtain Rabi frequencies $\Omega_{01}$ and $\Omega_{12}$ to accomplish the desired transformation of populations. 

\textit{Device}.--Here we use a  frequency-tunable superconducting Xmon qutrit \citep{Barends-Nature,Kelly-Nature,caoNC,ningPRL,Xukai-optica,YangNPJ} to test the above theory. The original Hamiltonian reads $H=\Omega_{01}e^{i\omega_{01}t}|0\rangle\langle 1|+\Omega_{12}e^{i\omega_{12}t}|1\rangle\langle 2|+{\rm H.c.}+\sum_ {i=0,1,2}\omega_i |i\rangle \langle i|$, where $\omega_i$ and $\omega_{01(12)}$ are the angular frequencies of energy level $|i\rangle $ and microwave pulses coupling $|0\rangle \leftrightarrow | 1\rangle\ (|1\rangle \leftrightarrow | 2\rangle)$, respectively. We adjust $\omega_{01}/2\pi=(\omega_1-\omega_0)/2\pi=5.9600$ GHz and $\omega_{12}/2\pi=(\omega_2-\omega_1)/2\pi= 5.7208$ GHz (in this experiment, frequency accuracy to 4 decimal places is mandatory) to ensure that the pulses can resonantly drive the transitions between adjacent energy levels. Note the fixed  frequency of the resonator (not used here) is 5.584 GHz \citep{caoNC,ningPRL,Xukai-optica,YangNPJ}, dynamically decoupled with the above system. 

\begin{figure}
\includegraphics[width=8.4cm]{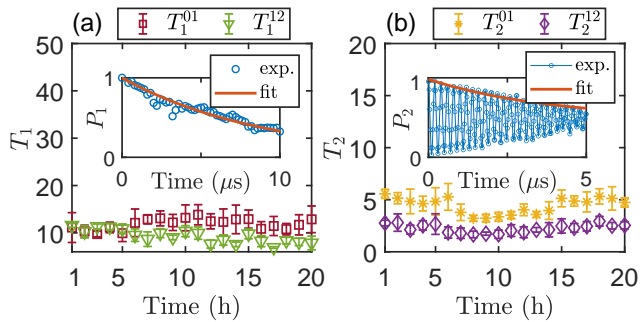}
\caption{(Color online) Measurement results of (a) relaxation time $T_1$ and (b) dephasing time $T_2$. The data are measured 12 groups per hour for a total of 20 hours. The insets, as  examples, are the fitting processes for $T_1$  and $T_2$ in (a) and (b), respectively.}
\label{T1T2}
\end{figure}

\begin{figure}
\includegraphics[width=8cm]{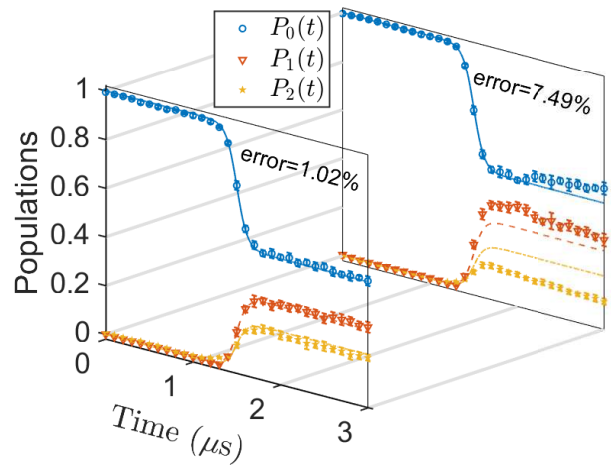}
\caption{(Color online) Experimental results of the control of populations $P_1(0)=0\to P_1(t_f)=0.3$ and $P_2(0)=0\to P_2(t_f)=0.2$. The experimental results of applying open-system and closed-system pulses are shown (with circles, triangles, and pentagrams) in the outer and inner layers, respectively. The ideal results of $P_0(t)$, $P_1(t)$, and $P_2(t)$ are indicated by solid, dashed and dotted dashed curves, respectively.}
\label{2layers}
\end{figure}

\begin{figure}
\includegraphics[width=8.4cm]{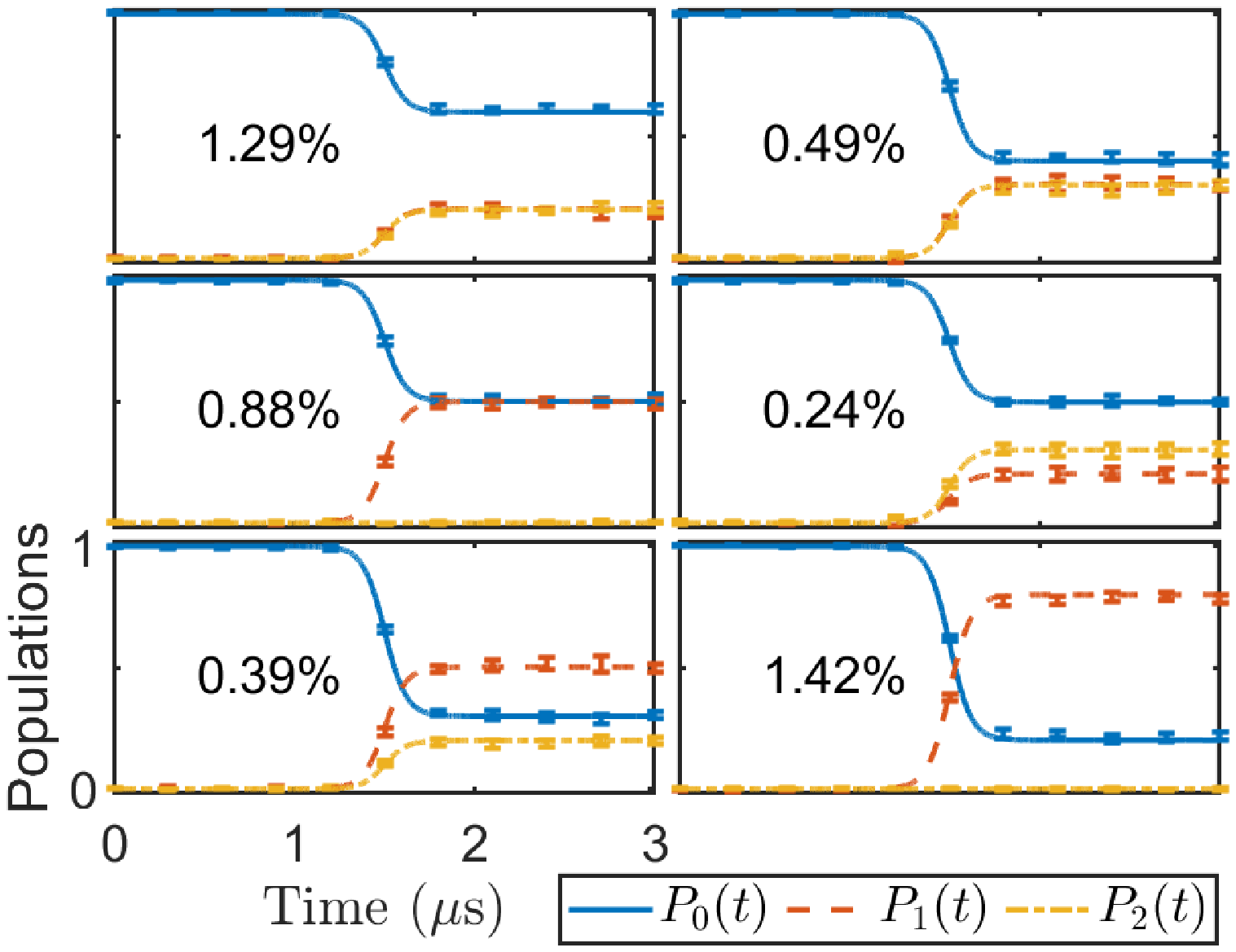}
\caption{(Color online) Six groups of experimental results (error bars) of the controls of populations. The ideal results of $P_0(t)$, $P_1(t)$, and $P_2(t)$ are represented by solid, dashed and dotted dashed curves, respectively. The error rates are shown in corresponding subgraphs.}
\label{tile6}
\end{figure}

\begin{figure}
\includegraphics[width=8.4cm]{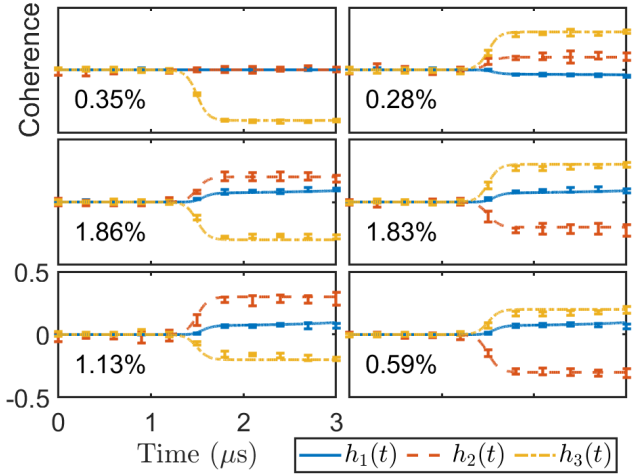}
\caption{(Color online) Six groups of experimental results (error bars) of the controls of coherence. The ideal results of $h_0(t)$, $h_1(t)$, and $h_2(t)$ are represented by solid, dashed and dotted dashed curves, respectively. The error rates are shown in corresponding subfigures.}
\label{tile6co}
\end{figure}

An essential process for this experiment is to precisely measure the coefficients of decoherence, $\gamma_k$ and $\Gamma_k$. According to the above form of Lindblad operators, one can deduce $\Gamma_1=1/T_1^{01}$, $\gamma_1=2/T_2^{01}-\Gamma_1$, $\Gamma_2=1/T_1^{12}$, and $\gamma_2=2/T_2^{12}-\Gamma_2-\Gamma_1-\gamma_1$. Here $T_1^{01(12)}$ and $T_2^{01(12)}$ are the energy relaxation and dephasing time measured between $|0\rangle$ and $|1\rangle$ ($|1\rangle$ and $|2\rangle$), shown in Figs. \ref{T1T2}(a) and \ref{T1T2}(b), respectively. From Fig. \ref{T1T2}, we find $T_1$ and $T_2$ fluctuate a lot during the measured period of 20 hours. In addition, several works \citep{martinis-spinecho,DDapply1,DDapply2,DDapply3,DDapply4,6hours,Guo-longT2} support that $T_2$ will change with the applying of microwave pulses (the spin-echo technique \citep{Echo} is a strong proof). Therefore the values of $T_1$ and $T_2$ we utilized to design pulses are a little different from those measured in Fig. \ref{T1T2}, specifically, $[T_1^{01},T_1^{12},T_{2}^{01},T_{2}^{12}]=[9.5, 4.6, 6, 1.9] \ \mu s$, which are fixed and utilized for all the experimental control (Figs. \ref{2layers}, \ref{tile6}, and \ref{tile6co}).

\begin{figure}
\includegraphics[width=8cm]{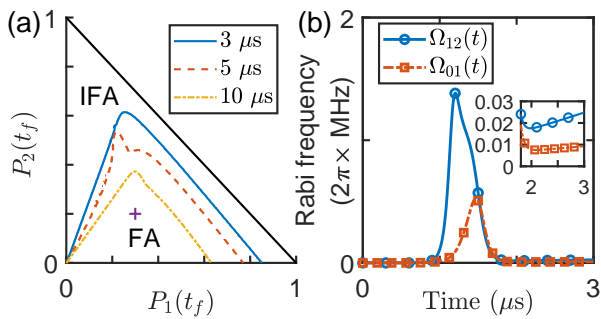}
\caption{ (Color online) (a) Feasible area (FA, surrounded by curves and x-axis) and infeasible area (IFA, the outside zone) of the populations control in the experiment when the evolution time is 3, 5, and 10 $\mu$s. Condition $P_1(t_f)+P_2(t_f)\le 1$ triangulates the boundary. (b) A sample point marked as purple plus sign in (a): Rabi frequencies for the control of populations $P_1(0)=0\to P_1(t_f)=0.3$ and $P_2(0)=0\to P_2(t_f)=0.2$.  The used parameters of decoherence are  $[T_1^{01},T_1^{12},T_{2}^{01},T_{2}^{12}]=[9.5, 4.6, 6, 1.9] \ \mu s$ and the inset shows the magnification picture of the Rabi frequencies.}
\label{areaandomega}
\end{figure}

\textit{Results of the populations control}.--Due to the choice of $f_1$ and $f_2$, there are infeasible zones of the populations transformation, shrinking as $t_f$ extending, shown in Fig. \ref{areaandomega}(a). 
%The present $f_1$ and $f_2$ are dominant in transforming the populations of between adjacent energy levels, however not good  at $|0\rangle \leftrightarrow | 2\rangle$. For such a two-photon transition, $f_1$ and $f_2$ can be designed as the superposition of error and Gaussian functions (detail in Supplemental material \citep{}). 
While the feasible area still takes a large part.  As an example, we utilize Eqs. (\ref{omega}, \ref{consteq}) to design $\Omega_{01}$ and $\Omega_{12}$ [see Fig. \ref{areaandomega}(b)] for the transformation of  populations $P_1(0)=0\to P_1(t_f)=0.3$ and $P_2(0)=0\to P_2(t_f)=0.2$. For the evolution time, we choose $t_f=3 \ \mu s$ to accumulate enough impact of the decoherence. The corresponding experimental results are shown in Fig. \ref{2layers}, with outer and inner layers indicating the results of applying open-system [see Eqs. (\ref{omega}, \ref{consteq})] and closed-system [see Eqs. (\ref{omega}, \ref{consteq}), preset $\gamma_k=\Gamma_k=0$] pulses, respectively. 
Intuitively, the populations are controlled more precisely in the outer layer, as compared to the inner one. More rigorously, we define a standard deviation to describe the error rate of the population control
\begin{eqnarray}\label{index}
{\rm error}=\sqrt{\sum_{i=0,1,2} [P_i(t_f)^{\rm ideal}-P_i(t_f)^{\rm exp.}]^2/3},
\end{eqnarray}
where $P_i(t_f)^{\rm ideal}$ and $P_i(t_f)^{\rm exp.}$ are the ideal and experimental values of the population in $|i\rangle$, respectively. In the outer layer of Fig. \ref{2layers}, the populations control achieves an error rate of 1.02\%, not small \citep{Standard} because we intentionally extended the evolution time (3 $\mu$s). In contrast, if the control is performed for such a long time by applying closed-system pulses, the error rate reaches 7.49\%. This stark difference in error demonstrates the effectiveness of the present control method. 

We also measure the results of 29 different controls of populations, 6 of them shown in Fig. \ref{tile6} and all displayed in the Supplemental Material \citep{Supp}. The error rates of the controls in Fig. \ref{tile6} are around 1\%, which we believe can be further improved with more stable superconducting qutrits \citep{Qiskit}.

Another exceptional physical phenomenon is that the populations seem to be frozen after 1.8 $\mu$s in the outer layers of Fig. \ref{2layers}. This freezing phenomenon of populations does not mean that the driving pulses $\Omega_{01}$ and $\Omega_{12}$ have stopped. On the contrary, it is the driving pulses [see the insets of Fig. \ref{areaandomega}(b)] we applied that caused this freezing phenomenon to occur. Specifically, such a freezing phenomenon arises from the interplay between the dynamics induced by the continuous microwave drives and the two decoherence channels characterized by $T_1$ and $T_2$ (both available for $|0\rangle \leftrightarrow | 1\rangle$  and $|1\rangle \leftrightarrow | 2\rangle$ transitions), similar to the cases for  generation of steady states in most open systems \citep{steady1,steady2,steady3,steady4,open-invariant,onlyT1,onlyT2}. But here it differs significantly in that both the energy relaxation and the dephasing are involved in the nonequilibrium dynamical processes and together help freeze the states of three-level systems, as compared to the previous ones which generally consider only one decoherence channel \citep{open-invariant,onlyT1,onlyT2}.


\textit{Results of the coherence control}. Since only two free variables, $\Omega_{01}$ and $\Omega_{12}$ in Eqs. (\ref{omega}, \ref{consteq}) can be controlled in the experiment, we can only control two parts of coherence. Here we choose $h_3$ and $h_2$ as an example for illuminating the coherence control. Corresponding variations of Eqs. (\ref{omega}, \ref{consteq}) and the forbidden zone of coherence control are shown in the Supplemental Material \citep{Supp}. According to the same preset of parameters and intermediate function $f$, we show 6 groups of results of the coherence control in Fig. \ref{tile6co} (all the 36 groups of results in the Supplemental Material \citep{Supp}). The error rate here is
\begin{eqnarray}\label{index2}
{\rm error}'=\sqrt{\sum_{m=2,3} [h_m(t_f)^{\rm ideal}-h_m(t_f)^{\rm exp.}]^2/2},
\end{eqnarray}
where $h_m(t_f)^{\rm ideal}$ and $h_m(t_f)^{\rm exp.}$ are the ideal and experimental values of coherence $h_m$, respectively. Figure \ref{tile6co} shows that $h_2$ and $h_3$ are controlled well and the values of $h_1$ are accurately predicted. Similarly, the microwaves protect the coherence (lasting 1.2 $\mu$s) from the influence of dephasing. If there are no microwaves applied, roughly,  after the same 1.2 $\mu$s, $h_3(1.2 \mu {\rm s})/h_3(0)\sim\exp(-1.2 \mu {\rm s}/T_{2}^{01})\sim 0.8$ and $h_2(1.2 \mu {\rm s})/h_2(0)\sim\exp(-1.2 \mu {\rm s}/T_{2}^{12})\sim 0.45$, the coherence will be seriously damaged. In contrast, the error rates are all within 2\% when we apply microwaves for the coherence control.
%These are interesting results because we succeeded in the precises dynamical control whose evolution time is close to relaxation and dephasing times, $T_1$ and $T_2$. In other words, the evolution time of such exact dynamical control maybe exceed 200 $\mu$s based on more advanced equipment \citep{Qiskit} ($T_1,T_2>200 \ \mu s$).

\begin{figure}
\includegraphics[width=8cm]{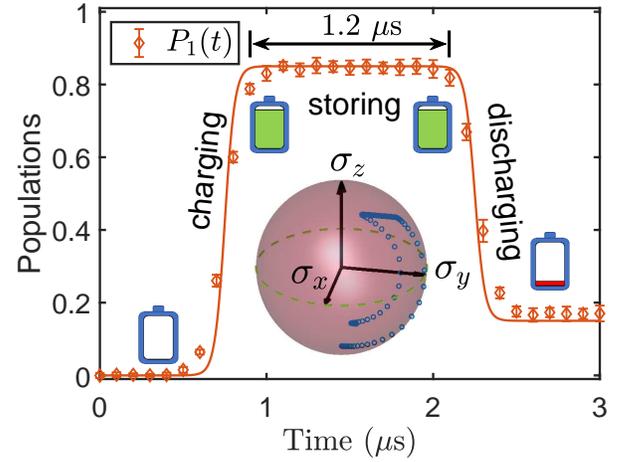}
\caption{(Color online) Experimental verification of the QB. We plot the experimental and ideal results by diamond points and solid curves, respectively. The inserted  circle depicts the Bloch sphere of the ideal evolution and the density of points represents the speed of evolution, one point per 15 ns. The parameters of decoherence are $[T_1^{01},T_{2}^{01}]=[9.5, 6]\ \mu$s.}
\label{QB}
\end{figure}

\textit{Application}.-- Quantum battery \citep{QB-initial,QB-PRL1,QB-PRL2,QB-PRL3,QB-NJP,QB-PRB,QB-PRA1,QB-PRA2,QB-PRA3,QB-PRE,QB-EPL}. The freezing phenomenon seems helpful to stabilize the energy of the QB for a fairly long time, especially in the energy-storing stage of the QB. For proving this, we design a pulse for a qutrit in the experiment to simulate the charging, storing, discharging processes of the QB (see Fig. \ref{QB}).  A considerably long period, about 1.2 $\mu$s, of energy [$\epsilon=(\omega_1-\omega_0) P_1$] stability was observed. This stability is hard to achieve by applying a pulse designed without considering decoherence because energy relaxation of the superconducting qutrit will exponentially drop the excitation down. Therefore, the present method may contribute to the construction of more stable QBs.

%It is worth mentioning that due to the  limitation of the acceptable area we can just maximally maintain excitation of 0.85 for 1.2 $\mu$s. %On the other side, if one only apply two $\pi$-pulses of 40 ns to charge and discharge the QB, and no pulse to store the energy, the excitation will be dissipated from 1 to 0.88 in 1.2 $\mu$s. 

\textit{Conclusion}.--We have proposed a method to control the dynamics of three-level open systems and realized it in the experiment with a platform of a superconducting Xmon qutrit \citep{Barends-Nature,Kelly-Nature,caoNC,ningPRL,Xukai-optica,YangNPJ}. %In the experiment of populations control, during a relatively long evolution time of 3 $\mu$s, we successfully reduce the magnitudes of error rates from 7\% to 1\%.
The populations and coherence could be singly controlled with error rates around 1\% under the influence of decoherence for a relatively long time, 3 $\mu$s, close to the dephasing time $T_2^{12}$ of the qutrit. In some situations where the control is more successful, the error rates can even be less than 0.3\%. We believe these error rates can be further reduced with more stable superconducting qutrits \citep{Qiskit}.

Additionally, an interesting freezing phenomenon of populations was observed. The designed microwave pulses just offset the impact of decoherence and visually freeze the populations. We then applied this phenomenon to make more stable prototypes of QBs, whose energy can hold 1.2 $\mu$s with only one charging. Moreover, this freezing phenomenon strongly proves that the Markovian master equation precisely describes the dynamics of three-level open systems, as demonstrated here with a superconducting Xmon qutrit that possesses the specially intrinsic decoherence rates. The present work provides a positive prospect of accurately realizing the dynamical control of three (or more)-level open systems by using the adjustable driving pulses in the Markovian environment.

Note that all the data points of experiments were averaged over 12000 times and the readout errors of the qutrit have been corrected. 
We thank  Ye-Hong Chen and Xin Wang in RIKEN for discussions. This work was supported by the National Natural Science Foundation of China (Grants No. 11874114, and No. 11875108), and the Natural Science Funds for Distinguished Young Scholar of Fujian Province under Grant 2020J06011, Project from Fuzhou University under Grant JG202001-2.

\bibliography{new5}
\end{document}

\title{Supplemental Material for “Demonstration of dynamical control of three-level open systems with a superconducting qutrit"}
\author{Ri-Hua Zheng}
\thanks{These two authors contributed equally to this work.}
\author{Wen Ning}
\thanks{These two authors contributed equally to this work.}
\author{Zhen-Biao Yang}\email{zbyang@fzu.edu.cn}
\author{Yan Xia}\email{xia-208@163.com}
\author{Shi-Biao Zheng}\email{t96034@fzu.edu.cn}
\affiliation{Fujian Key Laboratory of Quantum Information and Quantum Optics, College of Physics and Information Engineering, Fuzhou University, Fuzhou, Fujian 350108, China}
\maketitle
\section{experimental setup}
\begin{figure*}
\includegraphics[width=15cm]{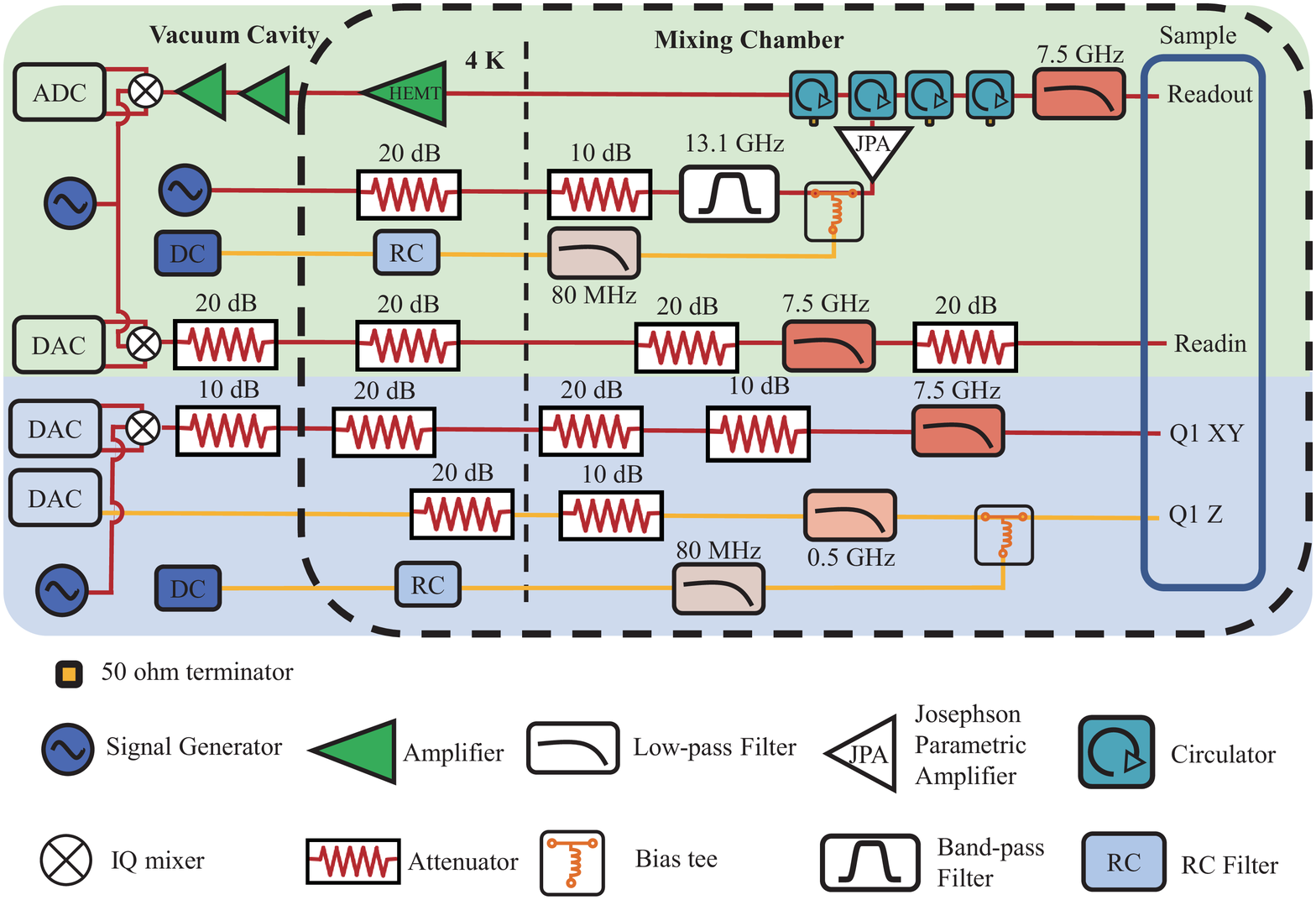}
\caption{(Color online) The wiring diagram of the experimental setup. Green top channel: readin and readout of the Xomn qutrit; blue bottom channel: XY and Z control of the Xomn qutrit. The detailed description is placed in the text.}
\label{Wiring}
\end{figure*}

The entire control and readout layout of the experiment is shown in Fig. \ref{Wiring}. Electronic devices for qutrit readout, Josephson parametric amplifier (JPA) control, and qutrit control are displayed from top to bottom.
The readin and XY control signals are the mixtures of low-frequency signals [from two independent digital-to-analog converter (DAC) channels I and Q] and high-frequency signals (from microwave sources) to achieve nanosecond fast tuning. On the other hand, the qutrit Z control signal is sent directly from the DAC without mixing, whose frequency can also be slowly tuned by the direct current (DC) bias line. 
Additionally, the output signal of the readout feeder is amplified by the JPA, high electron mobility transistor (HEMT), and room temperature amplifier, and then demodulated by the analog-to-digital converter (ADC). Four cryogenic unidirectional circulators are inserted between the JPA and the 4K HEMT to block the reflection and noise from outside. 
As for the JPA, it is pumped by an independent microwave signal source and a DC bias, which is converged by the bias tee. 
At different temperature stages of the dilution refrigerator, each control line is balanced with some attenuators and filters (intuitively shown in Fig. \ref{Wiring}) to prevent unwanted noise from affecting the equipment.
\section{qutrit readout and calibration}
\begin{figure}
\includegraphics[width=8cm]{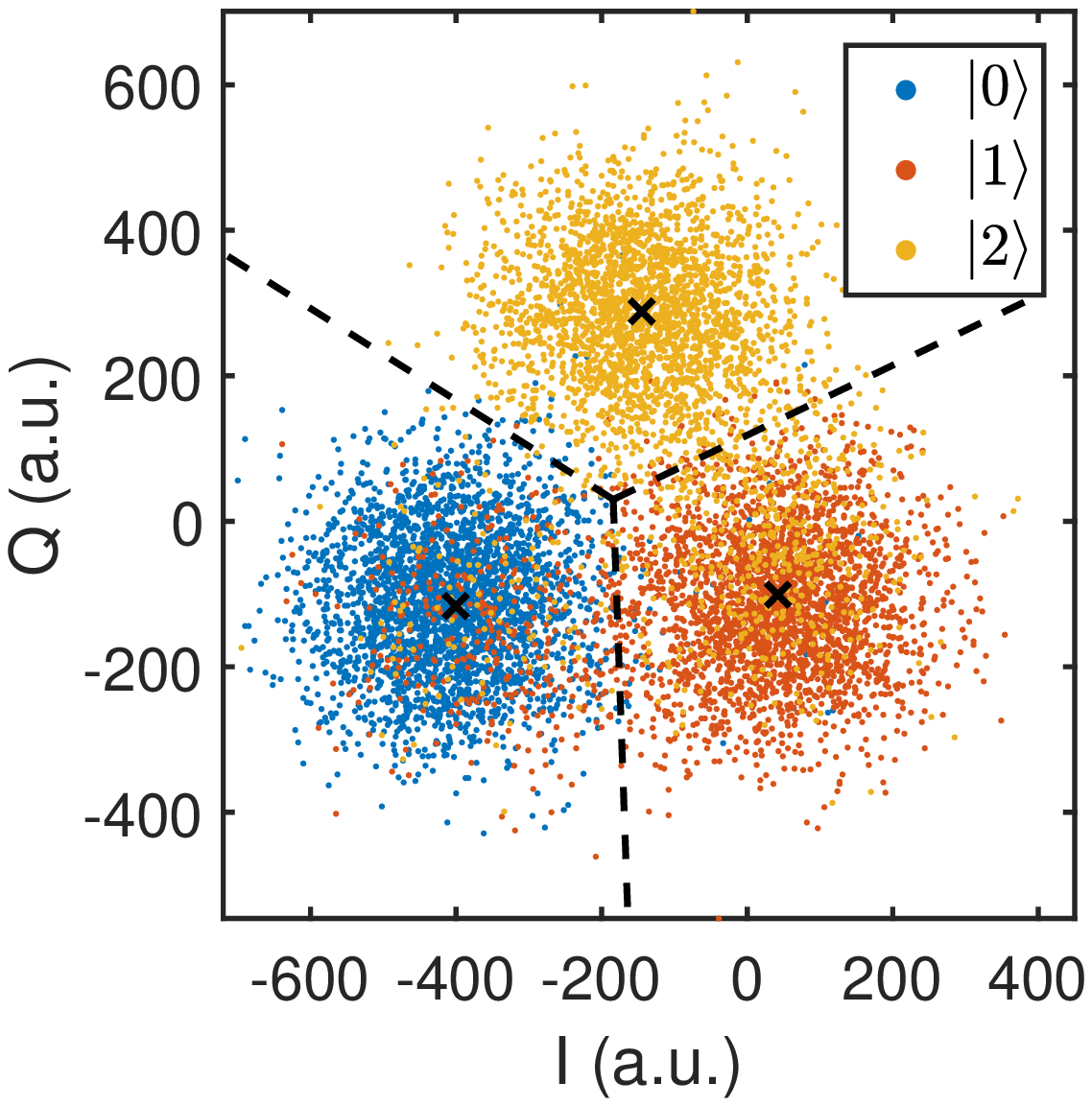}
\caption{(Color online) Results of the qutrit readout in the I-Q plane. Blue, orange and yellow points are the results of $|0\rangle$, $|1\rangle$, $|2\rangle$, respectively. The geometric centers of the three cluster points (3000 points per cluster) are marked by black cross signs.}
\label{IQraw}
\end{figure}

Here we show the results of the qutrit readout in Fig. \ref{IQraw}, where blue, orange, and yellow points represent the results of applying $I$, $X_{01}$, and $X_{01}X_{12}$ to the qutrit, i.e., depicting the results of preparing $|0\rangle$, $|1\rangle$, $|2\rangle$, respectively. 
The readout pulses are 1 $\mu$s-long and the repetition is 3000. 
It can be seen that these three states are clearly separated, though with several error transitions caused by the readout errors. Like that in Refs. \citep{cao-PRL,ning-PRL}, the calibration matrix can be defined as 
\begin{eqnarray}\label{calibration1}
F=\left(
\begin{array}{ccc}
F_{00} & F_{01} & F_{02} \\
F_{10} & F_{11} & F_{12} \\
F_{20} & F_{21} & F_{22} \\
\end{array}
\right),
\end{eqnarray}
where $F_{ii'}$ ($i,i'=0,1,2$) is the probability of measuring the qutrit in $|i\rangle$ when it is prepared in $|i'\rangle$. Such that the measured probability is given by $P_i^{\rm m}=\sum_{i'} F_{ii'}P_{i'}^{\rm c}$, with $P_{i'}^{\rm c}$ being the calibrated probability of $|i'\rangle$, viz.,
\begin{eqnarray}\label{calibration2}
\left(
\begin{array}{c}
P_{0}^{\rm c} \\
P_{1}^{\rm c} \\
P_{2}^{\rm c} \\
\end{array}
\right)=F^{-1}\left(
\begin{array}{c}
P_{0}^{\rm m} \\
P_{1}^{\rm m} \\
P_{2}^{\rm m} \\
\end{array}\right).
\end{eqnarray}
We can derive, for instance, $F_{01}=N_{\rm OL }/3000$, in which $N_{ \{\rm B,O,Y\} \rm \{L,R,T\}}$ is the number of \{blue, orange, yellow\} points in the \{left, right, top\} zones in Fig. \ref{IQraw}. In addition, we  repeat the measurement for 10 times to take the average values, yielding
\begin{eqnarray}\label{calibration3}
F=\left(
\begin{array}{ccc}
0.974 & 0.102 & 0.041 \\
0.017 & 0.885 & 0.141 \\
0.009 & 0.013 & 0.818 \\
\end{array}
\right).
\end{eqnarray}
Note that all the experimental data have been calibrated by the matrix in Eq. (\ref{calibration3}).

%\section{choice of the intermediate function $f$}
%For the intermediate function $f$ in the main text, $f=[1+e^{-a(t-t_f/2)}]^{-1}$, it is hard to realizes populations control when $P_1(t_f)=0$ and $P_2(t_f)\ne 0$ [i.e., $f_1\equiv0$ and $f_2=P_2(t_f)f$] because directly transition of $|0\rangle \leftrightarrow |2\rangle$ can not be achieved by $\Omega_{01}$ ($|0\rangle \leftrightarrow |1\rangle$) and $\Omega_{12}$ ($|1\rangle \leftrightarrow |2\rangle$). Alternatively, we can indirectly accomplish transition $|0\rangle \leftrightarrow |1\rangle \leftrightarrow |2\rangle$. This requires a non-zeros population $P_1(t)$ during the evolutionary process.

\section{variation of Eqs. (3, 4) in the main text for the coherence control}
Equations (3, 4) in the main text are used to control the populations, with designable $f_k$ ($k=1,2$) and iterative $h_{i+1}$ ($i=0,1,2$). If we alternatively aim to control two parts of the coherence (e.g., $h_2$ and $h_3$) of systems, we need to alter Eqs. (3, 4) in the main text as 
\begin{small}
\begin{eqnarray}\label{index}
&&\Omega_{01}=\frac{h_3 [(\gamma+\Gamma)+2\dot{h}_3]+h_2[ (\gamma_2+\Gamma_2)+2 \dot{h}_2]}{2 h_1 h_2-2 h_3 (2 f_1+f_2-1)}, \cr\cr
&&\Omega_{12}=\frac{ (2 f_1+f_2-1) [h_2(\gamma_2+\Gamma_2)+2 \dot{h}_2]+h_1 [h_3 (\gamma+\Gamma)+2\dot{h}_3]}{2 h_1 h_2-2 h_3 (2 f_1+f_2-1)}, \cr\cr
&&\dot{h}_1=\Omega_{12} (f_1-f_2)-\frac{1}{2} h_1 (\gamma+\Gamma+\gamma_2+\Gamma_2)-h_2\Omega_{01}, \cr\cr
&&\dot{f}_1=-\Gamma f_1+\Gamma_2 f_2+2h_3\Omega_{01}-2h_1\Omega_{12}, \cr\cr &&\dot{f}_2=-\Gamma_2f_2+2h_1\Omega_{12},
\end{eqnarray}
\end{small}
in which $h_{k+1}$ is designable and $f_k$ is iterative. Therefore one can design suitable $h_2$ and $h_3$ to accomplish desired coherence control.

\section{Feasible and infeasible areas of the coherence control}
\begin{figure}
\includegraphics[width=8cm]{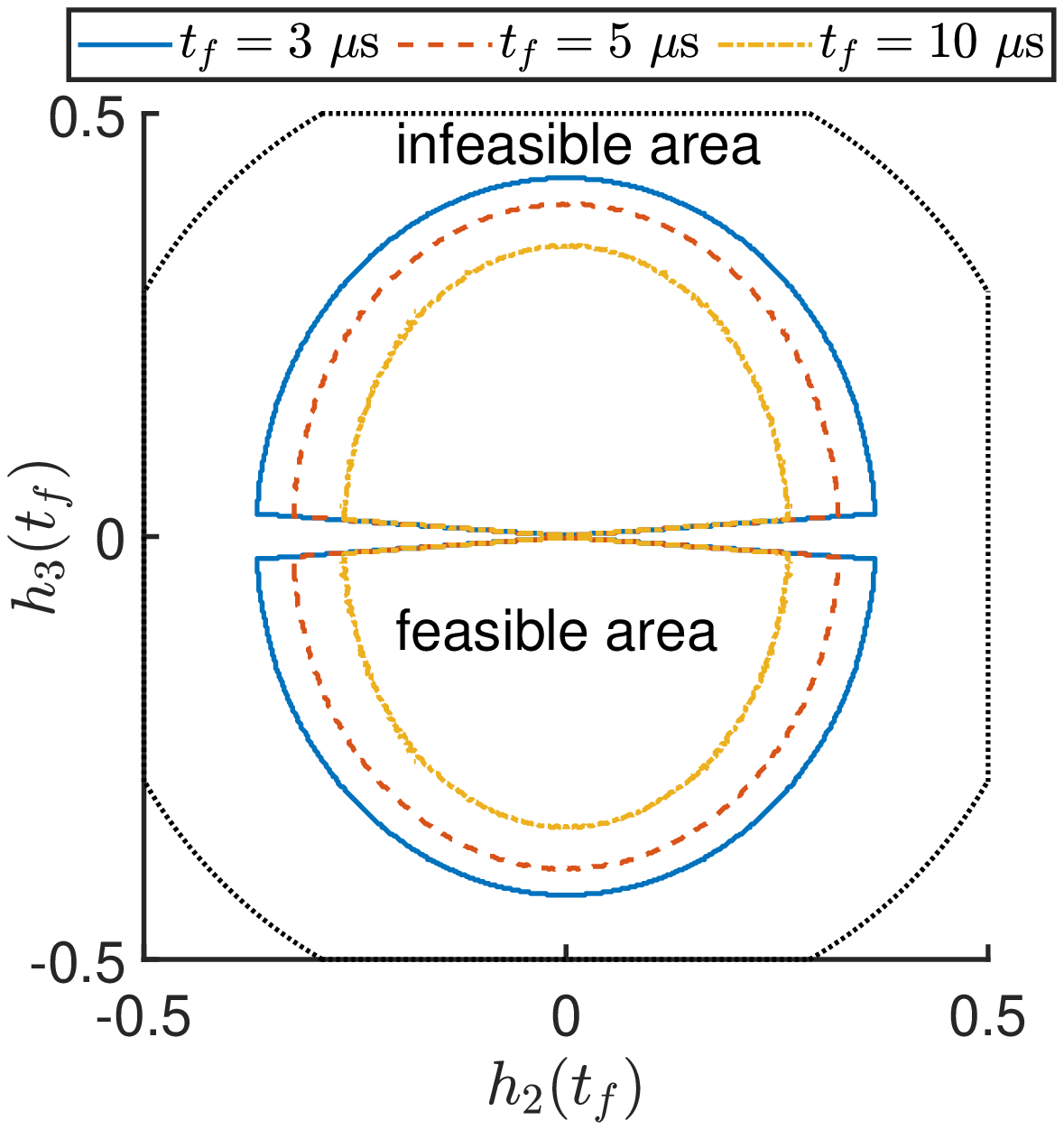}
\caption{(Color online) Feasible area (surrounded by solid, dashed, and dotted dashed curves) and infeasible area (the outside zone surrounded by dotted curves) of the coherence control in the experiment when the evolution time is  3, 5, and 10 $\mu$s.}
\label{IFAco}
\end{figure}

For the control of $h_2$ and $h_3$, similarly, there are feasible area (FA) and infeasible area (IFA), which mainly depend on the intermediate function $f$ and decoherence. For $f=[1+e^{-a(t-t_f/2)}]^{-1}$, we plot FA and IFA of the coherence control in Fig. \ref{IFAco}. The total area is constructed by conditions $[h_2,h_3]\le0.5$ and $h_2^2+h_3^2\le [1-f_1^2-f_2^2-(1-f_1-f_2)^2-2h_1^2]/2\le1/3$.

\section{Tomography of the coherence control}
The populations of three-level systems can be directly measured by the readout cavity. (In the experiments, only diagonal elements of the density matrix can be directly measured.) While the coherence can be indirectly measured by a simplified tomography, including 4 steps as follows, 

(i) $U_1=I$,
\begin{eqnarray}\label{U1}
\rho_1=U_1\rho U_1^\dag=\left(
\begin{array}{ccc}
f_2 & -ih_1 & h_2 \\
ih_1 & f_1 & -ih_3 \\
h_2 & ih_3 & 1-f_1-f_2 \\
\end{array}
\right),
\end{eqnarray}
(ii) $U_2=(X/2)_{01}$,
\begin{small}
\begin{eqnarray}\label{U2}
\rho_2&=&U_2\rho U_2^\dag\cr\cr 
&=&\left(
\begin{array}{ccc}
f_2 & -\frac{i (h_1-h_2)}{\sqrt{2}} & \frac{h_1+h_2}{\sqrt{2}} \\
\frac{i (h_1-h_2)}{\sqrt{2}} & -\frac{f_2}{2}+h_3+\frac{1}{2} & \frac{1}{2} i (2 f_1+f_2-1) \\
\frac{h_1+h_2}{\sqrt{2}} & -\frac{1}{2} i (2 f_1+f_2-1) & \frac{1}{2} (-f_2-2 h_3+1) \\
\end{array}
\right),
\cr  &
\end{eqnarray}
\end{small}
(iii) $U_3=(X/2)_{12}$,
\begin{small}
\begin{eqnarray}\label{U3}
\rho_3&=&U_3\rho U_3^\dag\cr\cr&=&\left(
\begin{array}{ccc}
\frac{f_1+f_2}{2}+h_1& \frac{-i}{2}(f_1-f_2)& \frac{1}{\sqrt{2}}(h_2-h_3) \\
\frac{i}{2}(f_1-f_2) & \frac{1}{2}(f_1+f_2-2h_1) & \frac{-i}{\sqrt{2}}(h_2+h_3) \\
\frac{1}{\sqrt{2}}(h_2-h_3) &\frac{i}{\sqrt{2}}(h_2+h_3) & 1-f_1-f_2 \\ 
\end{array}
\right),\cr &
\end{eqnarray}
\end{small}

(iv) $U_4=U_3U_2$,
\begin{small}
\begin{widetext}
\begin{eqnarray}\label{U4}
\rho_4&=&U_4\rho U_4^\dag\cr\cr&=&\
\left(
\begin{array}{ccc}
\frac{1}{4} \left(f_2+2 \sqrt{2} h_1-2 \sqrt{2} h_2+2 h_3+1\right) & \frac{1}{4} i (3 f_2-2 h_3-1) & \frac{2 f_1+f_2+\sqrt{2} h_1+\sqrt{2} h_2-1}{2 \sqrt{2}} \\
-\frac{1}{4} i (3 f_2-2 h_3-1) & \frac{1}{4} \left(f_2-2 \sqrt{2} h_1+2 \sqrt{2} h_2+2 h_3+1\right) & -\frac{i \left(-2 f_1-f_2+\sqrt{2} h_1+\sqrt{2} h_2+1\right)}{2 \sqrt{2}} \\
\frac{2 f_1+f_2+\sqrt{2} h_1+\sqrt{2} h_2-1}{2 \sqrt{2}} & \frac{i \left(-2 f_1-f_2+\sqrt{2} h_1+\sqrt{2} h_2+1\right)}{2 \sqrt{2}} & \frac{1}{2} (-f_2-2 h_3+1) \\
\end{array}
\right),
\end{eqnarray}
\end{widetext}
\end{small}
where $I$ represents identity gate and $(X/2)_{01(12)}$ denotes a $\pi/2$ rotation over the X axis of Bloch sphere in basis \{$|0\rangle$,$|1\rangle$\} (\{$|1\rangle$,$|2\rangle$\}). For all the experimental data points of the coherence (i.e., $h_{i+1}$), we measure three diagonal elements of $\rho_{p}$ ($p=1,2,3,4$), signed as $\rho_{p}^{(ii)}$ ($i=0,1,2$), and further deduce $h_{i+1}$, i,e.,
\begin{eqnarray}\label{tomoh2h3}
h_1&=&\rho_{3}^{(22)}-\frac{\rho_{1}^{(22)}+\rho_{1}^{(11)}}{2}, \ \ h_3=\rho_{2}^{(11)}-\frac{1-\rho_{1}^{(22)}}{2}, \cr\cr
h_2&=&-\frac{\rho_{1}^{(22)}-2\sqrt{2}h_1+2h_3-4\rho_{1}^{(11)}+1}{2\sqrt{2}}.
\end{eqnarray}
By now, a simplified tomography for reading out coherence $h_{i+1}$ of three-level systems is completed.

\section{Supplementary experimental data for the dynamical control}
\begin{figure*}
\includegraphics[width=17.5cm]{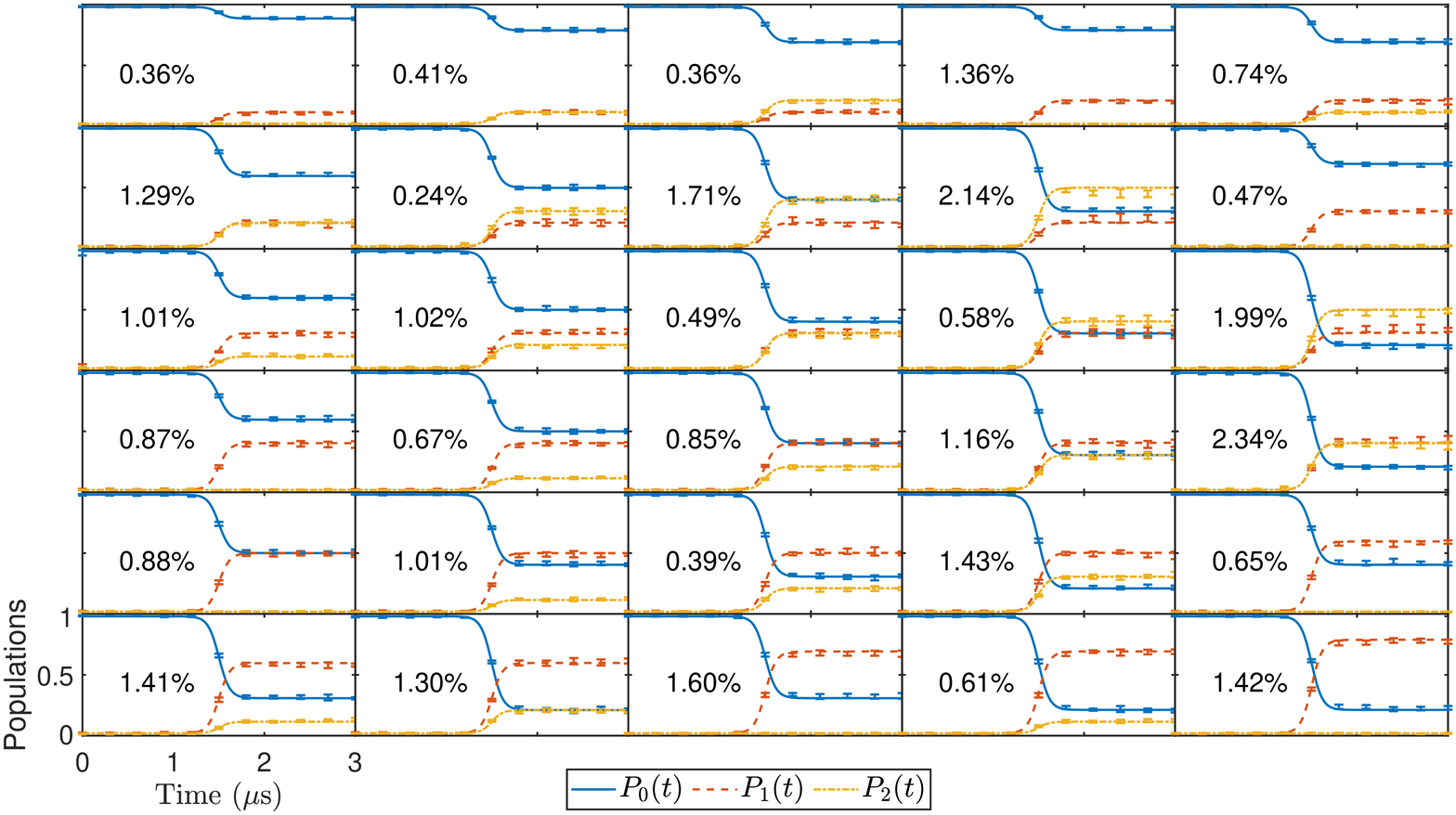}
\caption{(Color online) Experimental results (error bars) of the controls of populations. The ideal results of $P_0(t)$, $P_1(t)$, and $P_2(t)$ are represented by solid, dashed and dotted dashed curves, respectively. The error rates are shown in corresponding subgraphs.}
\label{tile}
\end{figure*}

\begin{figure*}
\includegraphics[width=17.5cm]{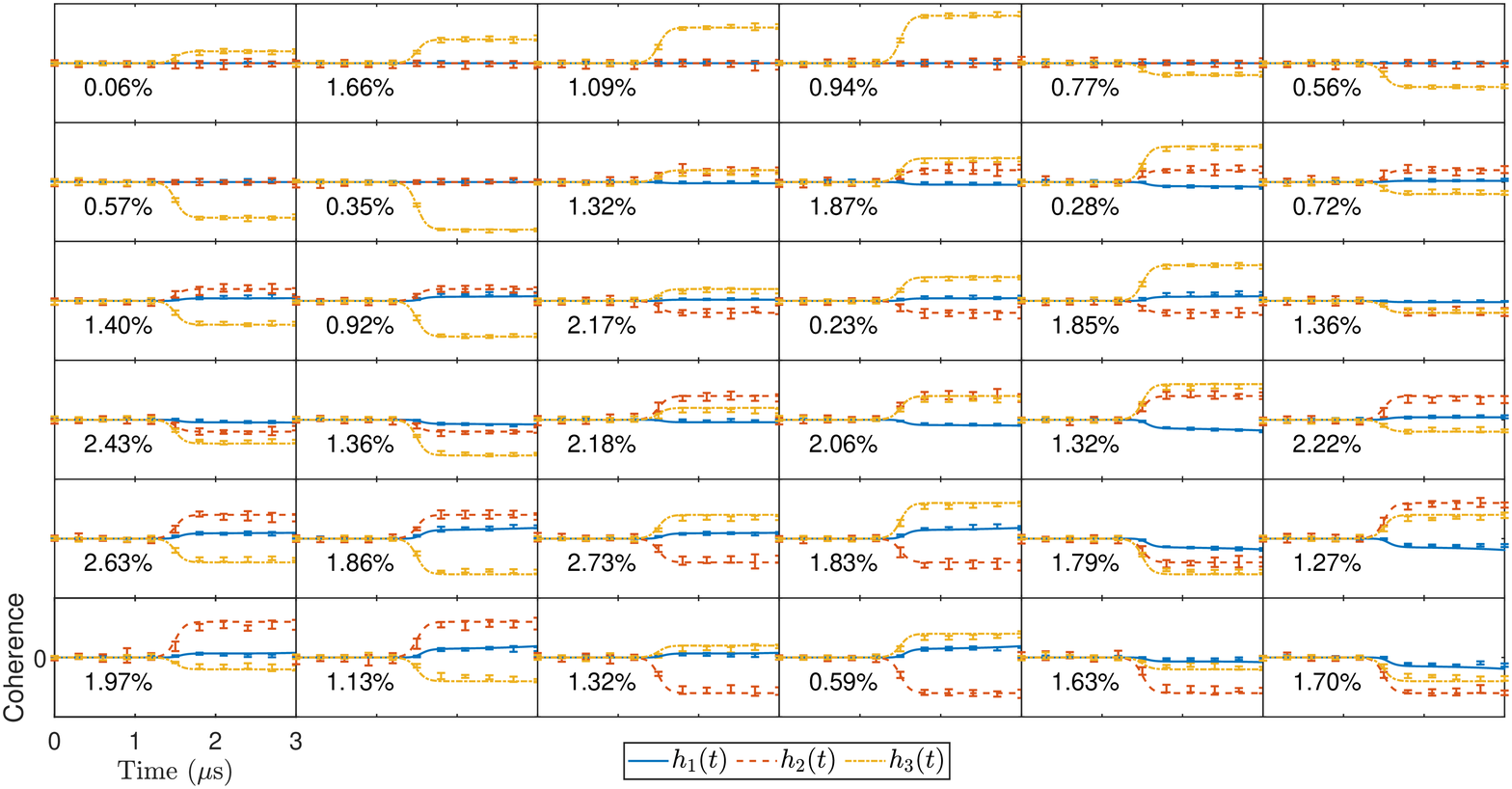}
\caption{(Color online) Experimental results (error bars) of the controls of coherence. The ideal results of $h_0(t)$, $h_1(t)$, and $h_2(t)$ are represented by solid, dashed and dotted dashed curves, respectively. The error rates are shown in corresponding subfigures.}
\label{tile36}
\end{figure*}

Here we show 30 and 36 groups of experimental data for the populations and coherence control in Figs. \ref{tile} and \ref{tile36}, respectively, whose average error rates are 1.02\% and 1.39\%, respectively, demonstrating the feasibility of the dynamical control in three-level open systems.

\bibliography{supp}